\documentclass{svjour3}                     
\smartqed  
\usepackage{graphicx}
\usepackage{natbib}
\usepackage{comment}
\journalname{EA}
\graphicspath{{Figures/}}

\begin{document}

\title{Mapping Distortion of Detectors in UVIT Onboard AstroSat Observatory}


\author{V Girish         \and
        Shyam N. Tandon \and S. Sriram \and Amit Kumar\and Joe Postma
}


\institute{V Girish \at
              Space Astronomy Group, ISRO Satellite Center \\
              HAL Airport Road, 
              Bangalore - 560017, INDIA\\
              \email{giri@isac.gov.in}           
           \and
           Shyam N. Tandon \at
		   Inter University Centre for Astronomy and Astrophysics,
		   Ganeshkhind, Pune, 411007, India\\
		   Indian Institute of Astrophysics, 
		   Koramangala II block, Bangalore - 560034, INDIA
		   \and
           S. Sriram, Amit Kumar \at
		   Indian Institute of Astrophysics,
		   Koramangala II block, Bangalore - 560034, INDIA
		   \and
		   Joe Postma\\
		   University of Calgary, Canada
}

\date{Received: date / Accepted: date}

\maketitle

\begin{abstract}

	Ultraviolet Imaging Telescope (UVIT) is one of the payloads onboard
AstroSat, India's first multi-wavelength Astronomy mission. UVIT is
primarily designed to make high resolution images in wide field, in  three
wavelength channels simultaneously: FUV (130 - 180 nm), NUV (200 - 300 nm)
and VIS (320 - 550 nm). The intensified imagers used in UVIT suffer from
distortions, and a correction is necessary for these to achieve good
astrometry. In this article we describe the methodology and calculations
used to estimate the distortions in ground calibrations.

\end{abstract}


\section{ Introduction }

Ultraviolet  Imaging Telescope (UVIT; \citealt{amit12}) is one of the
payloads on board the Indian Astronomy Satellite AstroSat
\citep{agarwal05,singh14}.  UVIT is primarily designed to make high spatial
resolution ($<$ 1.8$''$) images over a field of ${\sim}$28$'$, simultaneously in
three channels: FUV (130-180 nm), NUV (200-300 nm), and VIS (320-550 nm).
Soft X-ray telescope (SXT; \citealt{archna10}) and Scanning Sky monitor
(SSM; \citealt{rama16}) are the other two imaging payloads on-board
AstroSat capable of imaging the sky in soft X-ray bands with arc-minute
resolution.

Due to the inherent design of the intensified imagers, involving a fiber
taper for reducing the plate-scale, the final image suffers from
distortions which interfere with good astrometry.  Therefore, a correction
for the distortions is necessary for good astrometry.  For example, by
comparing the SDSS source catalogs with GALEX detections, an absolute
astrometric precision of 0.49$''$ rms and 0.59$''$ rms for NUV and FUV
respectively was achieved for
GR2/GR3 data \citep{morrissey07} which is further improved to 0.32$''$ rms
amd 0.34$''$ respectively in the latest GR6 data \citep{galex}. 

The distortions in UVIT are primarily in the CPU (Camera Proximity Unit
containing the Detector Module, shown in Figure~\ref{uvit-dm} and some
electronics mounted on the focal plane of the detector).  The distortions
in the CPU are mainly believed to be due to the imperfections in the Fiber-
taper construction. In principle it is best to calibrate these distortions of
the full optical chain in the orbit. However, it is difficult to find a
dense astrometric field for ultraviolet, and only low frequency spatial distortions
might be measured in the orbit. Therefore we plan to derive distortions in
two steps:

\begin{figure}
	\includegraphics[width=4in]{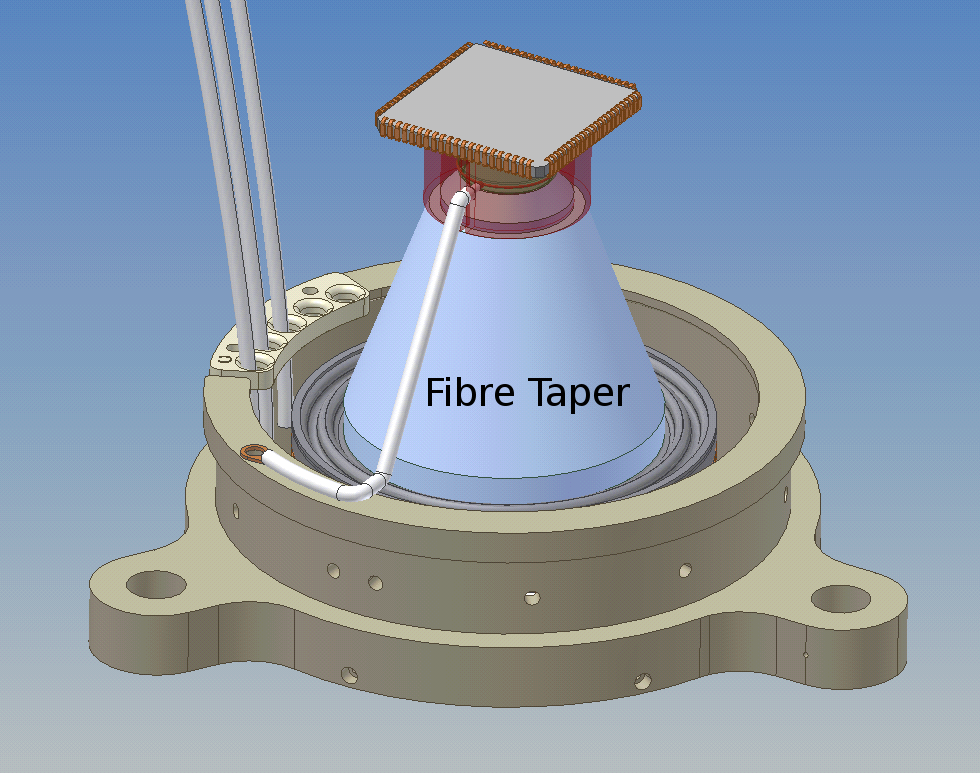}
	\caption{Schematic of UVIT detector module is shown. The fibre taper is
	indicated, with the CMOS imager at the top.}
	\label{uvit-dm}
\end{figure}

\begin{enumerate}
\item
Distortion in each CPU is measured by imaging a simple geometrical pattern,
e.g. a set of parallel lines or a grid of pin holes etc.  

\item
An astrometric field in the orbit  exposed during the AstroSat performance and
verification (PV) phase  to confirm the distortions estimated on ground.  
\end{enumerate}

Here we discuss the experimental procedure and analysis of observations
used for estimating distortions in the CPUs on ground using a grid of holes
drilled in some geometrical pattern. The distortion values obtained at
discrete positions are then extended to all the pixel positions by
interpolating the values. The pixel positions which fall outside of the
grid pattern (edges of the detector) are filled by extra-polating the
values.

In the second part of the calibrations which will be published elsewhere,
we will compare the ground calibration results with the on-board
observations of selected open clusters to check the efficacy of applying
distortion values to the observed fields and improvement in the astrometry.

\section{Distortions in CPUs }

For estimating the distortions in the CPUs, images of a grid are taken by
the CPUs; the optical arrangement to obtain these images is shown in Figure
2.  These images are called as ``IIA images'' from now onwards.  The grid is
made by drilling pinholes, of size ${\sim}$0.04 mm, with laser beam on a substrate of
Invar. The pinholes are in a square pattern, covering a circle of ${\sim}$40 mm
diameter, as shown in Fig~\ref{fig-mask}. The task of estimating the distortions is done
in two parts.  In the First part, the ``exact positions'' of centres of the
pinholes are estimated, and in the second part positions of the centres in
the images taken with the CPUs are compared with the ``exact positions''.

\begin{figure}
	\includegraphics[width=4.2in]{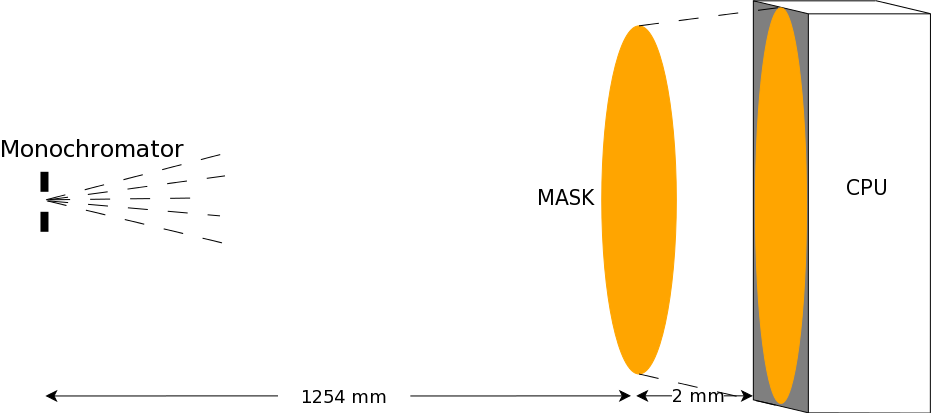}
	\caption{Schematic of the optical arrangement used for estimation of
	distortions of CPUs. }
	\label{cpu_optical}
\end{figure}

\begin{figure}
	\includegraphics[width=5in]{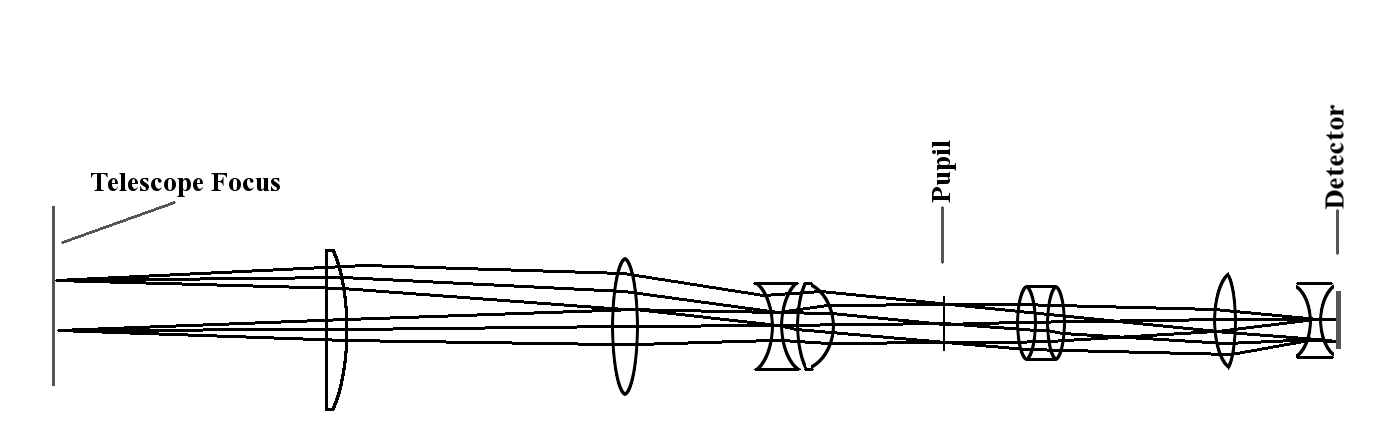}
	\label{fig-ifosc}
	\caption{ Optical layout of IUCAA Faint Object Spectrometer and Camera
	(based on \citealt{ranjan02}).}
\end{figure}

\begin{figure}[h]
\includegraphics[width=3.85in]{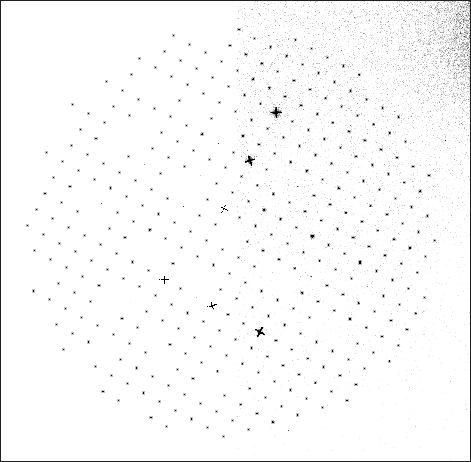}
\caption{Image of the grid as taken in R-filter with IFOSC of IUCAA telescope.
}
\label{fig-mask}
\end{figure}


\subsection{ Positions of the Centres of the Pinholes obtained with IFOSC}
\label{sec-ifosc}

\subsection{ Images Taken with IFOSC }



To obtain centroids of the pinholes, the grid is imaged by the
high quality reducer of IUCAA Faint Object Spectrograph \& Camera (IFOSC;
\citealt{ranjan02}), at
Ineter University Center for Astronomy and Astrophysics (IUCAA), Pune in, R
and V bands. Schematic of IFOSC is shown in Fig~\ref{fig-ifosc}. The grid
was placed in the slit-holder located at the ``Telescope Focus''. The images were
recorded, with a reduction by a factor ${\sim}$2.2, on a CCD with 2168 pixel ${\times}$
2048 pixel of 13.5 $\mathrm{{\mu} m}$. Two sets of data were taken in each R and V band. In the second set, the grid
was rotated by ${\sim}$180 degree
with reference to the first set. The observations were repeated for three
exposures, 5s, 10s and 15s each to obtain optimum dynamic range and to
avoid saturation of pixels as the holes are of different sizes as seen in
Fig~\ref{fig-mask}.
For obtaining the centers of the pinholes, we use only IFOSC
images with 10s and 15s integrations.

\noindent


The pinhole positions of individual images are
obtained with the help of \emph{daophot} \citep{stetson87} routine provided
with 'ASTROLIB' \citep{landsman93}.  A two dimensional Gaussian is fitted
to each pinhole to obtain the centers.   For unique identification of
the individual holes, a set of larger holes is drilled in a pattern of `L'
as can be seen in Fig~\ref{fig-mask}.
Each `larger hole' consists of multiple pinholes in a ``+'' pattern.
The ``FIND'' routine
detects these larger pinholes as multiple closely
spaced sources and these closely spaced groups of positions are removed
manually. The positions corresponding to other
larger and un-even shaped pinhole images are also removed.

\subsection{ Combining the Positions from Different Images} 

The IFOSC images of the grid  can have mainly two defects:

\begin{itemize}
\item

{\bf Wedge effect:} There could be some angle between the focal plane and
the CCD and this would give unequal gains at different positions ( the
optics is NOT telecentric). This can be corrected by suitably combining
images from the two sets.

\item

{\bf Distortion due to optics:} data on distortion are obtained by
simulating the optics in Zemax; a polynomial fit with, even powers of
radius,  $b *r^2 + c*r^4$, is made to these data as the optics of IFOSC
is centro-symmetric. Positions of the
pinholes in the images are corrected for this.

\end{itemize}

As mentioned earlier, there are two sets of images taken for two
different orientations of the grid, which differed by a rotation of ~180
degrees. An average of the positions in these two sets would correct for
any distortion due to wedge effect. In order to take such average, it is
essential to convert all the positions to the same frame of reference by
allowing for any deviations in the angle of rotation from 180 degrees and
for any shifts of centre of the grid. This process is described below. 

\begin{quote}

Assuming that there is no warping in the images, the transformation will be
a linear transformation and involves only rotation and scaling with offset.
For a scale `g',  rotation `${\theta}$' and offsets $x_0$ \& $y_0$, the translated
positions (x$_i'$, y$_i'$) corresponding to the position (x$_i$, y$_i$) can
be written as,

\begin{displaymath}
x'_i = g[x_i \cos{\theta}- y_i \sin{\theta}]+x_0 
\end{displaymath}
\begin{equation}
y'_i = g[x_i \sin{\theta}+ y_i \cos{\theta}]+y_0
\label{eqn-rotat}
\end{equation}

The positions viz., $x_0, y_0, x_i, y_i, x'_i, y'_i$ are in
units of pixels and the angle is in radians. 
To determine the rotation angle and the scale, we use the method of
minimising sum of the squares of deviations,


\begin{equation}
f(g,{\theta}, x_0, y_0) = {\sum}_i [(X_i - x'_i)^2 + (Y_i - y'_i)^2] 
\label{eqn-dist}
\end{equation}

where, $X_i$ and $Y_i$ are the observed positions in the reference image
(for IUCAA IFOSC data, the image observed in R-band and integrated for 15 s
is taken as the main image or reference image), and $x'_i$ and $y'_i$ are
the transformed (zoomed, rotated with offset) positions of the comparison
frame/image given by Eqn.\ref{eqn-rotat}.

For ${\theta}{\sim}0$, $\sin{\theta}= {\theta}$ and $\cos{\theta}= 1$.
Also g=1 for images taken with same telescope, hence
Equation~\ref{eqn-dist} will yield, 

\begin{equation}
f(g, {\theta}, x_0, y_0) = {\sum}_i \{[X_i - g(x_i  - y_i {\theta})-x_0]^2 +
						[Y_i - g(x_i {\theta}+ y_i)-y_0]^2\} 
\end{equation}

where, ($x_i, y_i$) are the original positions in the comparison image.

The parameters of fit are found by
equating the four partial derivatives $\frac{{\partial} f}{{\partial} {\theta}}$, $\frac{{\partial} f}{{\partial} g}$,
$\frac{{\partial} f}{{\partial} x_0}$ and $\frac{{\partial} f}{{\partial} y_0}$ to zero and solving the linear
equations. We note that before application of the above procedure, the
coordinate system for the comparison image is rotated/shifted manually so
that positions of the centres nearly match in the two images. The averaged
positions are given by,

$$ X'_i = X_i - \frac{1}{(J+1)}{\sum}_{j=1}^{j=J} F_j(x_i) $$
\begin{equation}
	Y'_i = Y_i - \frac{1}{(J+1)} {\sum}_{j=1}^{j=J} F_j(y_i)
\label{eqn-diff}
\end{equation}

where, j = 1,2.. corresponds to different corrections obtained by comparing
the reference frame (j=0, R-filter, 15~s, 0 deg rotation) with other
frames. 
F($x_i$) = $[X_i - g(x_i  - y_i {\theta})-x_0]$ and F($y_i$) = $[Y_i -
g(x_i {\theta}+ y_i)-y_0]$.  F(x$_i$) and F(y$_i$) actually gives the errors
in the position X$_i$ and Y$_i$.
\end{quote}

\begin{figure}
\includegraphics[width=4.5in]{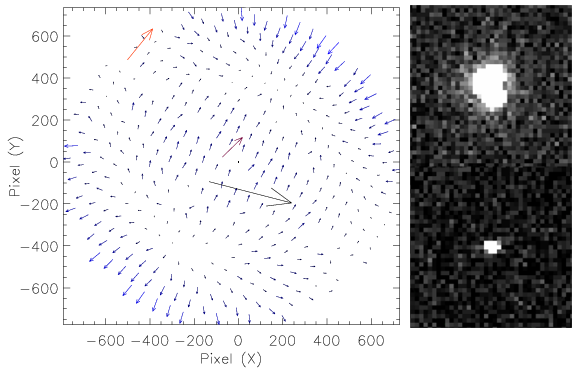}
\caption{Plot of difference between the center positions of IFOSC reference image
(R-filter, 15 s integration, 0 deg rotation) and
comparison image (R-band, 15~s, 180 deg grid rotation) after fitting.
Close-up view of the two positions showing large deviation
are shown on the RHS. The top one has a deformed shape and the second
largest deviation (bottom image) falls very close to the edge.}
\label{fig-fitFull}
\end{figure}

The difference of the centers of the reference and the matched image is
plotted in Figure~\ref{fig-fitFull}. The plot shows that the difference in
the positions of the centers of pinholes after rotation is less than 0.25
pixel except for three pinhole positions.
A plot of the  difference vectors
is also shown in Figure~\ref{fig-fitFull}.

{Figure~\ref{fig-fitFull} suggests  that only three points have large
differences as compared to the average difference of other points. A close
inspection of these points (as shown in the RHS of Fig~\ref{fig-fitFull})
shows that: in the first pinhole position (1), the pinhole is not circular.
The second pinhole (2) falls very close to the edge.  Hence, the three points
are removed from the main and compared images and subjected to least square
fit again.  Figure~\ref{fig-fit} shows the plot of the difference vector
after subtracting the three largest error vectors from the plot shown in
Fig~\ref{fig-fitFull}.
The exercise of obtaining the differences is repeated for all the remaining
six images and an average difference map is obtained by averaging the seven
difference maps thus obtained. The average difference map is plotted in RHS
of Figure~\ref{fig-fit}. The mean and rms of this average
difference map are 0.05 pixel and 0.04 pixel respectively. The maximum
difference is 0.23 pixels close to the bottom center shown in blue color in
the figure.

\begin{figure}
\includegraphics[width=2.40in]{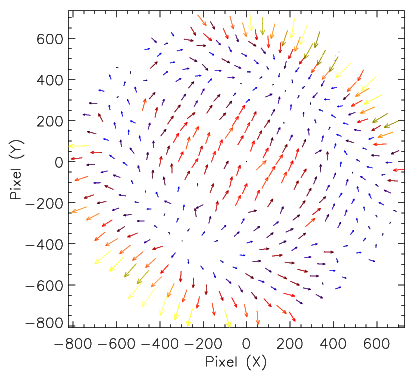}
\includegraphics[width=2.40in]{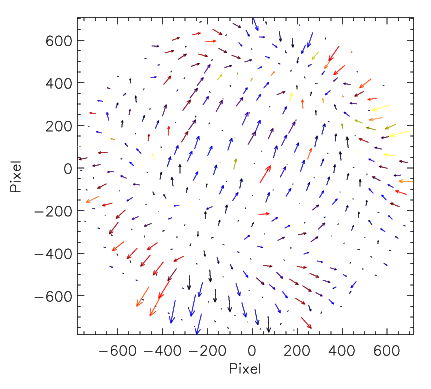}
\caption{ Similar to Figure~\ref{fig-fitFull} after the second iteration
after eliminating the three
pinholes which show large deviations.
RHS is the difference plot of average of
six IFOSC images obtained by comparing with R-filter image taken with zero degree rotation. The maximum is 0.23 pixel which is near the middle of the bottom
of the figure shown in blue color.}
\label{fig-fit}
\end{figure}

\subsection{ Correcting for Distortions in IFOSC images }

As mentioned earlier, the images of the grid were taken by a camera called
IFOSC and this camera has some distortion. Simulations with Zemax have been
fitted to derive the distortion as,

\begin{equation}
	r' = \frac{2.244}{5}r + \frac{0.010}{625}r^2 +
	\frac{0.016}{(625*625)}r^4
\label{eqn-rad1}
\end{equation}

where, ``r$'$'' is the radius from centre of the CCD in ``mm'' and ``r'' is
the radius on the object plane in ``mm''.

From Eq~\ref{eqn-rad1}, we can write the approximation (which corrects for
the distortion and the magnification/gain) as:

\begin{equation}
	\frac{2.244}{5}r = r' - \left(\frac{5}{2.244}r'\right)^2 ~
	\left(\frac{0.010}{625}\right) - \left(\frac{5}{2.244}r'\right)^4 
	\left(\frac{0.016}{(625*625)}\right)
\label{eqn-rad2}
\end{equation}

The position coordinates ($X_p', Y_p'$) in units of CCD pixels can be
transformed to distance ($X' and Y'$) in units of ``mm'' as,

\begin{equation}
  X' = (Xp' - 1084) {\times}0.0135 
  \label{eqn-rad5}
\end{equation}
\begin{equation}
  Y' = (Yp' - 1024) {\times}0.0135 
  \label{eqn-rad6}
\end{equation}
\begin{equation}
  r' = \sqrt{(X')^2 + (Y')^2)}
  \label{eqn-rad7}
\end{equation}

For the two co-ordinates, one can also write,

\begin{equation}
  X {\times} 2.244/5 = X' {\times}r''/r' 
  \label{eqn-rad3}
\end{equation}

\begin{equation}
  Y {\times}2.244/5 = Y' {\times}r''/r' 
  \label{eqn-rad4}
\end{equation}
   where  $r'' = 2.244/5r$.

Equations~\ref{eqn-rad3} and \ref{eqn-rad4} give the 
corrected coordinates in the object plane in ``mm''.

{\bf Procedure :}

The procedure for estimating the radial distortion is given below.

\begin{figure}
\includegraphics[width=2.5in]{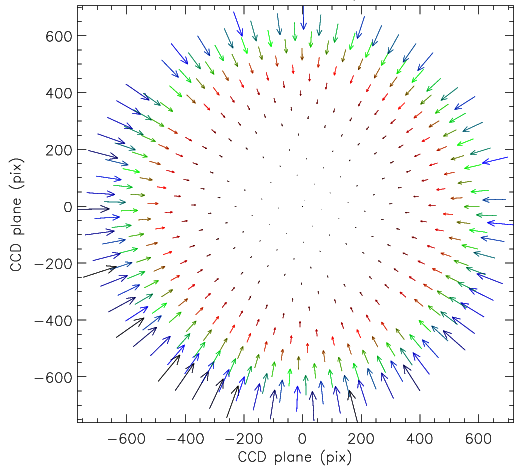}
\caption{Difference vector plot of estimated radial distortions after
correcting for the magnification.}
\label{fig-rad}
\end{figure}

\begin{enumerate}
\item
Positions of the pinholes are converted from pixels of the CCD
to mm according to Eqn~\ref{eqn-rad5} and
\ref{eqn-rad6}.

\item
Radial distance of the pinholes from the center  `r$'$' is calculated
according to Eqn~\ref{eqn-rad7}.
\item

Estimate r$''$.

\item

Corrected positions are calculated according to eqn 3 \& 4.
\item

The difference between X and X$'$*5/2.244 gives the radial distortion 
in the X-direction. Similarly the distortions along Y-direction are
estimated. The value ``5/2.244'' is introduced to account for the
magnification.
\end{enumerate}

The difference between the distortion corrected positions X \& Y and the
CCD positions X$'$ \& Y$'$ are obtained as below. 
	   $${\delta}X_i = X_i -X_i'*5/2.244 \,\, \mathrm{and}\,\, {\delta}Y_i = Y_i - Y_i'*5/2.244$$
A difference vector plot of the radial distortions (rx, ry) corrected for
magnification is shown in Figure~\ref{fig-rad}.  The maximum radial
distortion corresponds to 48.95 ${\mu}$m at position (-14.8 mm, -18.5 mm). The
maximum distortion is shown as a black colored vector in
figure~\ref{fig-rad}.  
After correction for the radial distortion the positions are given as,
$$ X_i = X'_i + {\delta}X_i, \,\,\,\,  Y_i = Y'_i + {\delta}Y_i$$ 

\section {Estimation of Distortion  in the CPUs}
\label{sec-cpu}

To estimate distortions in the CPUs, IIA images of the grid taken with the CPUs
are analysed to find positions of the pinholes. These positions are
compared with the final positions derived using IFOSC camera as
discussed in Subsection
\ref{sec-ifosc}. The positions obtained from IFOSC camera will be
called ``Reference Positions'' from here afterwards. As the magnification and
orientation of the images taken with CPUs is not predetermined, these are
taken as unknowns to derived by comparing the positions in the IIA images with
the ``Reference Positions''. Any differences between the positions found,
after correcting for the magnification and orientation, and the ``Reference
Positions'' are taken as distortions. In order to find any overall
distortion pattern, analysis is also made to find any elliptical
distortion. The procedures for finding the distortion are described below.

\subsection{ Procedure for Finding CPU Distortion }

The procedure for obtaining the magnification and the orientation of the
images is similar to that described for comparing images of the grid taken
with IFOSC (see $\S$~\ref{sec-ifosc}).
Thus, the following steps are implemented
\begin{enumerate}
	\item
		The X and Y positions of the grid pattern in the  IIA images are 
		obtained by using daophot and are then refined using a 2d Gaussian
		fit.

		The units are 1/8 of a pixel (called sub-pixel) of Star250.
	\item
		The multiple detections and positions corresponding to
		distorted/large holes are removed manually.
	\item

		The  ``Reference Positions'' are transformed for gain, rotation,
		and shift manually so as to obtain a close fit with the positions
		in the IIA image.
		
	\item
		The final values of gain, rotation, and shift for the ``Reference
		Positions'' are obtained by minimising sum of the squares of
		deviations in positions of the pinholes. 

	\item
		The difference between the IIA image and IFOSC positions (corrected 
		for the final rotation, gain and offset) gives the distortions.
		$${\Delta}X_i = imX_i - (IFOSCx_i - {\theta}{\times}IFOSCy_i)*g - X0 $$
		$${\Delta}Y_i = imY_i - ({\theta}{\times}IFOSCx_i + IFOSCy_i)*g - Y0 $$
		where, imX$_i$ and imY$_i$ are the X \& Y positions of the IIA
		image in sub-pixel. IFOSCx$_i$ \& IFOSCy$_i$ are the pre-rotated 
		IFOSC  positions.
		
	\item
		The exercise is repeated for images with multiple orientations of
		the grid (rotation angles), with the constraint that the gain is
		identical for all the images.
\end{enumerate}

The estimated distortion for the three grid positions taken with the FUV
are plotted as vector plots in Figure~\ref{fig-fuvsingle}.
The fit parameters obtained for different grid positions of the FUV images
are listed in Table~\ref{tab-fuv}.

	\begin{table}[h]
	\caption{The parameters for fitting the positions with ``Reference
	positions'', for three orientations of the grid, are shown for FUV, NUV and
	VIS CPUs.}
	\label{tab-fuv}
\begin{center}
	\begin{tabular}{lcrcccc}
		\hline\hline
		& g		&  \multicolumn{1}{c}{${\theta}$} & X0	 & Y0 & Max & mean ${\pm}$ rms \\
		& 	&  \multicolumn{1}{c}{(degree)} 	 & (sub-pixel)	 &
		(sub-pixel) & (sub-pixel) & sub-pixel \\\hline
		FUV 0 & 3.1511242 $^{\ddagger}$	& 179.670 & 88.51   & -46.62 & 10.92 & 4.36 ${\pm}$ 2.05\\
	  FUV 180 & 3.1511242 $~$			& 359.938 & -141.62 &	1.55 & 10.55 & 4.42 ${\pm}$ 2.00 \\
	  FUV 270 & 3.1511242 $^{\ddagger}$	& 269.692 &  14.85  &  96.24 & 10.72 & 4.39 ${\pm}$ 1.94 \\ &
	  \\

		NUV 0 & 3.147702 $^{\ddagger}$ & 179.539 &  85.08  & -54.40 & 15.36 & 6.40 ${\pm}$ 2.89\\
	  NUV 180 & 3.147702 	$~$		 & 359.353 & -153.55 &  18.98 & 15.38 & 6.17 ${\pm}$ 2.67 \\
	  NUV 270 & 3.147702 $^{\ddagger}$ & 269.823 &  15.28  &  87.16 & 19.92 & 6.76 ${\pm}$ 3.10
	  \\ & \\

		VIS 0 & 3.17179  $^{\ddagger}$ & 269.573 & -37.85 & 156.47 & 34.01 & 18.24 ${\pm}$ 6.13 \\
	  VIS 180 & 3.17179  $^{\ddagger}$ & 359.351 & -168.49&  24.52 & 35.25 & 17.91 ${\pm}$ 6.80 \\
	  VIS 270 & 3.17179        $~$   & 269.388 &  0.80  & 161.18 & 34.90 & 18.34 ${\pm}$ 6.62 \\ \hline
	\end{tabular}
	\end{center}
\end{table}

\begin{figure}
	\label{fig-fuvsingle}
	\includegraphics[width=4.5in]{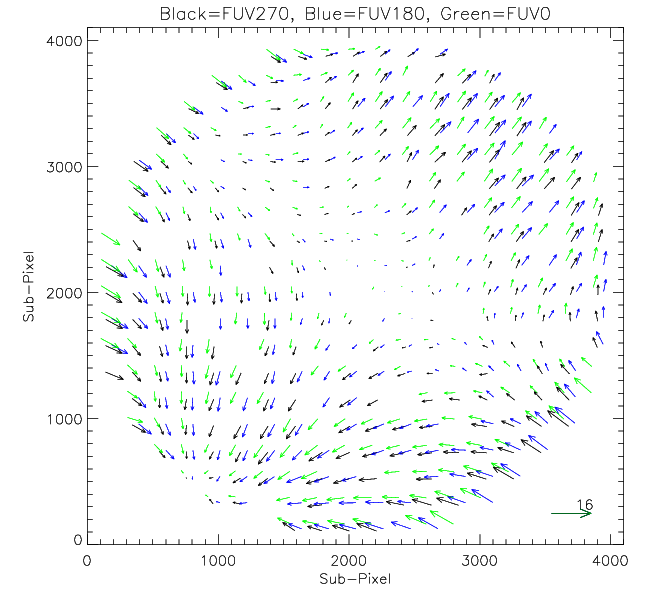}
	\caption{Vector plots of estimated distortion of the FUV detector.
	The three colors correspond to the three views of
	the grid: 0 deg (green), 180 deg (blue) and 270 deg (black) respectively.
	The size of the vector at the top left correspond to 16 sub-pixel.}
	\label{fig-fuvsingle}
\end{figure}

\begin{figure}
	\includegraphics[width=2.5in]{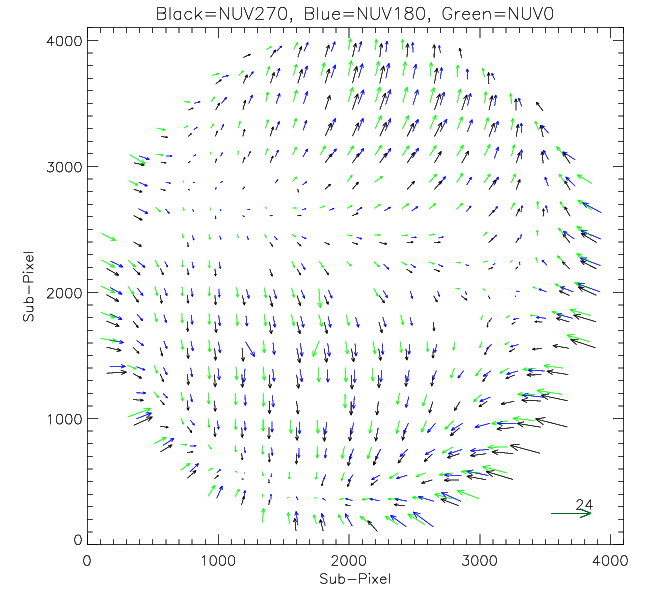}
	\includegraphics[width=2.5in]{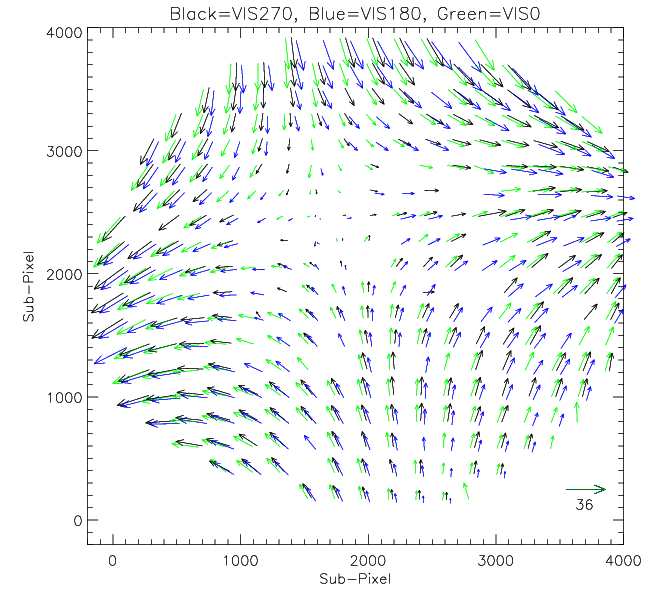}
	\caption{Similar to Figure~\ref{fig-fuvsingle} for NUV (L) and VIS (R) channels.
	The largest vector at the bottom right correspond to 24.0 and 36
	sub-pixels respectively.}
	\label{fig-nuvsingle}
\end{figure}


\subsection{Fit for Elliptical Distortion}

The figures (\ref{fig-fuvsingle}) \& (\ref{fig-nuvsingle}) showing the error vectors in the preceding subsection indicate
presence of some elliptical distortion. The procedure for estimating this
distortion is described below. 

\noindent
a) The distortion is simulated by assuming unequal gain along two
orthogonal directions.\\
b) The directions can have any angle with reference to the axes of the
image.\\
c) The ``Reference Positions'' are first transformed to a frame with axes at a
defined angle.\\
d) The next transformation is applied for two unequal gains along the two
axes.\\
e) Now the positions are transformed back to a frame with the original
direction of axes, with a shift of origin.\\
f)   The parameters of transformations are found by minimising sum of the
squares of differences in positions with reference to those in the image
with CPU\\
g) Step ``d)'' is tried for many angles in step ``c)'' and angle for the best
fit is chosen.\\

The above steps are captured by the following equations:

\begin{eqnarray}
\nonumber X''&  =  & a \cos (T) [X \cos (T)- Y \sin (T)] +  
b \sin (T)[X \sin (T) + Y \cos (T)] - X_0 \\
& = & X[a \cos^2(T) + b \sin^2(T)] + 
Y [ b \sin(T) \cos(T) - a \sin(T) \cos(T)] - X_0 \\
\nonumber Y''&  =  & b \cos (T) [X \sin (T)+ Y \cos (T)] -  
a \sin (T)[X \cos (T) - Y \sin (T)] - Y_0 \\
& = & Y[b \cos^2(T) + a \sin^2(T)] - 
X [ a \sin(T) \cos(T) - b \sin(T) \cos(T)] - Y_0
\end{eqnarray}

where `X' and `Y' are the best fit reference positions obtained from steps
3 \& 4 of $\S 3.1$, `T' is the angle of directions along which the gains
 are `a' and `b' respectively, and X$_0$ and Y$_0$ give the shift of
 origin.

The sum of the square of the distances (d) between the image
co-ordinates (imX$_i$, imY$_i$) and the modified reference frame
(X$''_i$, Y$''_i$) can be written as,

\begin{equation}
d^2 = {\sum}[(imX_i - X''_i)^2 + (imY_i - Y''_i)^2]
\end{equation}

The parameters `a', `b', `X0' and `Y0' are found by equating the 
partial derivatives	$\frac{{\partial}d}{{\partial}a}$, $\frac{{\partial}d}{{\partial}b}$, $\frac{{\partial}d}{{\partial}X0}$ and
$\frac{{\partial}d}{{\partial}Y0}$  to zero and solving the resulting linear equations.

This process is repeated for several values of `T', and the best value of
`T' (as inferred by sum of the squares of deviations) is selected.

The ``Remainders'' (the vectors given by $Fx_i$ and $Fy_i$)
for the three images, taken with FUV detector, are plotted in
Figure~\ref{fig-fuvdbl}.
The parameters of the fits  obtained for transformation with the
corresponding best value of `T' are listed in Table~\ref{tab-dblgain}.
The procedure is repeated for the remaining two detectors (NUV and VIS) and
the results are presented in Table~\ref{tab-dblgain} and in
Figure~\ref{fig-nuvdbl} respectively.

\begin{figure}
	\includegraphics[width=4.5in]{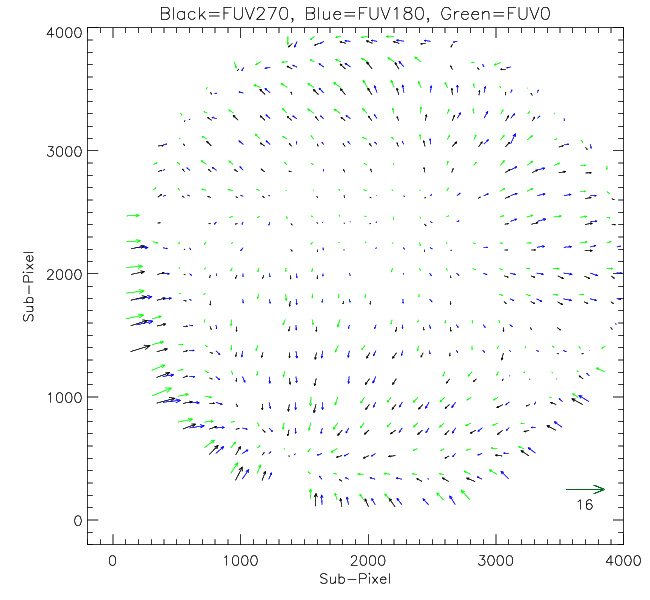}
	\caption{Vector plots of estimated distortion of the FUV detector with
	elliptical gain fits. The three colors corresponding to the three views of
	the grid: 0 deg (green), 180 deg (blue) and 270 deg (black) respectively.
	The size of the vectors at the top right correspond to the highest
	distortion of 16.0 sub-pixel.}
	\label{fig-fuvdbl}
\end{figure}

\begin{figure}
	\includegraphics[width=2.5in]{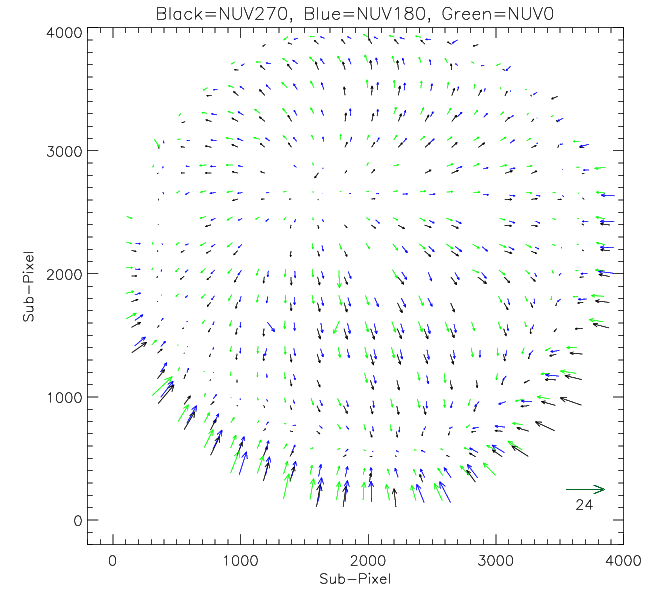}
	\includegraphics[width=2.5in]{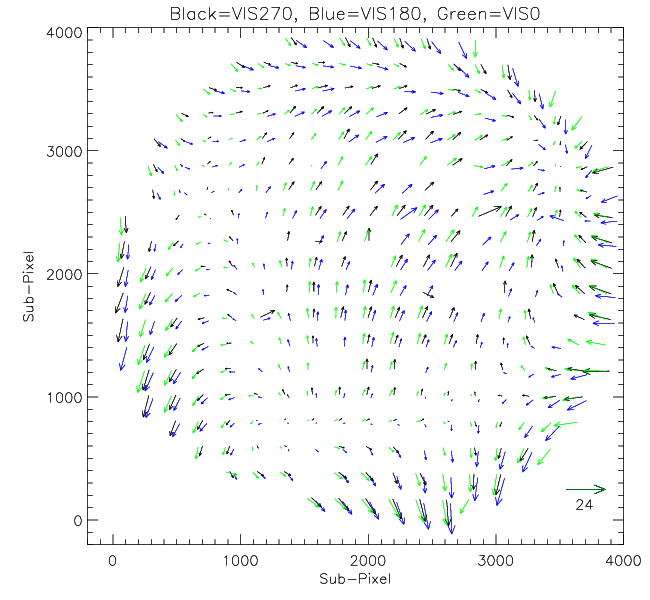}
	\caption{Similar to Figure~\ref{fig-fuvdbl} for NUV (L) and VIS (R)
	channels. The
	largest vector at the bottom right correspond to 24.0  sub-pixel.}
	\label{fig-nuvdbl}
\end{figure}

	\begin{table}[h]

	\caption{Double gain (elliptical) fit parameters obtained by
	minimizing the difference between positions of IIA images and
	average IFOSC image positions shown in Fig.~\ref{fig-fuvdbl} and
	\ref{fig-nuvdbl}.  The values marked with ${\ddagger}$ are fixed.}

	\label{tab-dblgain}
\begin{center}
	\begin{tabular}{lccccccc}
		\hline\hline
& $T$	  & a	 					& b 	 & X$_0$	&  Y$_0$ & Max & mean ${\pm}$ rms\\
&   & 	 	 				&   	 & sub-pix & sub-pix  &
sub-pix & sub-pix \\\hline
FUV 0   & 30$^o$ & 0.99713   $^{\ddagger}$ & 1.00279$^{\ddagger}$ & -0.07 &   0.07 & 8.76 &2.18 ${\pm}$ 1.32  \\
FUV 180 &    & 0.99713   $~$         & 1.00279$~$	      &  0.23 &  -0.03  & 7.29 &2.19 ${\pm}$ 1.17 \\
FUV 270 &    & 0.99713   $^{\ddagger}$ & 1.00279$^{\ddagger}$ &  0.00 &  0.00  & 8.48 &2.25 ${\pm}$ 1.35 \\~\\

NUV 0   & 20$^o$ & 0.99698$^{\ddagger}$    & 1.00364$^{\ddagger}$ & -0.19 &-0.04  & 17.16 & 4.55 ${\pm}$ 2.78 \\
NUV 180 &    & 0.99698$~$            & 1.00364$~$		  & -0.03 & 0.01  & 15.46 & 4.25 ${\pm}$ 2.80 \\
NUV 270 &    & 0.99698$^{\ddagger}$    & 1.00364$^{\ddagger}$ &  0.00 & 0.00  & 15.64 & 4.57 ${\pm}$ 2.77 \\~\\

VIS 0   & 80$^o$ & 0.98785$^{\ddagger}$    & 1.01286$^{\ddagger}$ &  0.40 & -0.27 & 16.11 & 5.99 ${\pm}$ 3.10\\
VIS 180 &    & 0.98785$~$            & 1.01286$~$		  &  0.86 & -0.85 & 19.33 & 6.34 ${\pm}$ 3.37 \\
VIS 270 &    & 0.98785$^{\ddagger}$    & 1.01286$^{\ddagger}$ &  0.67 &  1.17 & 16.53 & 6.22 ${\pm}$ 3.15 \\
\hline
\end{tabular}
\end{center}
\end{table}

\section{ Interpolation of distortion values }
 
To obtain the distortion values at each CCD250 pixel position, the discrete
distortion values obtained in Section~\ref{sec-cpu} are interpolated using the
utility interpolate.griddata available under python Scipy package
\citep{jones01}. The procedure followed for individual channels are as
follows using FUV channel as example:

\begin{figure}[htp]
	\includegraphics[width=4.6in]{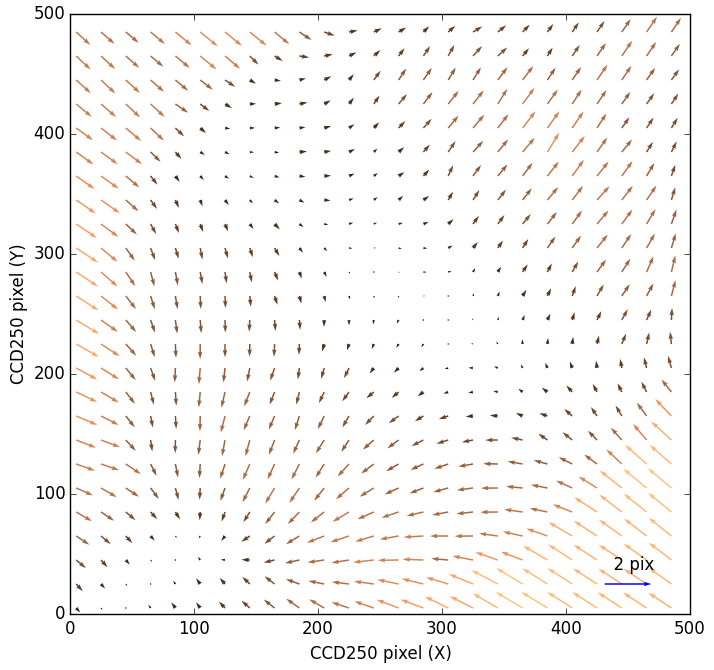}
	\includegraphics[width=2.3in]{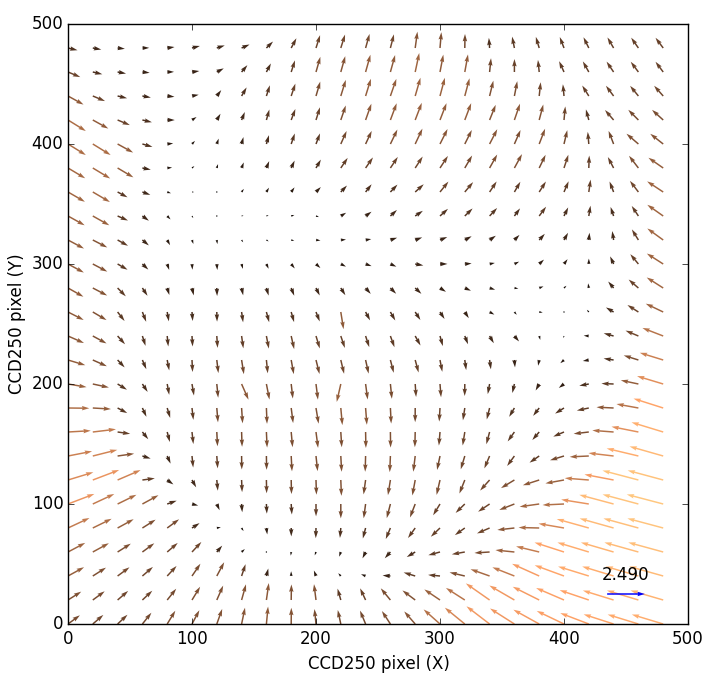}
	\includegraphics[width=2.3in]{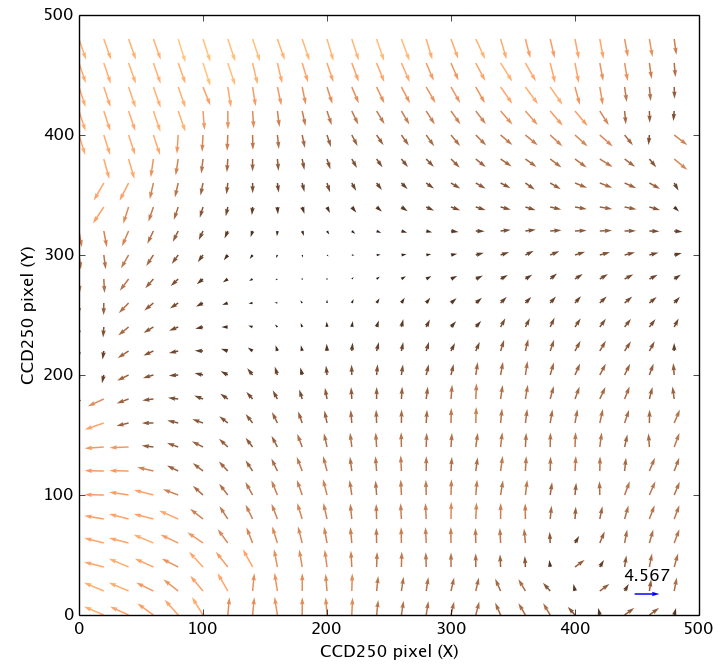}
	\caption{ Interpolated distortion values of FUV (top), NUV (bottom
	left) and VIS (bottom right) channels. Only every 20th positions are 
	plotted for clarity.}
	\label{fig-fuvinterpol}
\end{figure}

\begin{enumerate}
	\item
	Combine the X and Y-distortion values corresponding to the three
	rotations separately to increase the density of values.
\item
	Subject the combined (grid) values to interpolation using the method
	``linear''. This fills all the values that can be interpolated with the
	values and the values beyond the edges with NAN (Not a number).
\item
	Re-do the interpolation of the combined values from point 1 using the
	method ``nearest''. 
\item
	Replace all the NAN values from (2) with the corresponding values from
	(3) above.
\end{enumerate}

The resulting 512 x 512 array is the final distortion matrix corresponding
to X and Y-directions of FUV channel as difference vectors is shown in
Figure~\ref{fig-fuvinterpol}. The above procedure is repeated for other two
channels to obtain the distortion maps of all the three channels. The
resulting interpolated maps are also shown in Figure~\ref{fig-fuvinterpol}.
Note that for clarity, only every twentieth position is shown in the figure.

The 512 x 512 values corresponding to the X and Y-directions are
written as individual tables of a fits file. The three fits files
thus forms the UVIT distortion matrices and used for correcting the
observed UVIT fileds.  An analytic solution for the distortion map was not
tried as  a suitable analytic form which could 
reproduce the sharp changes near edges of the field is not available.
%

%
%
%

%
%

\section{ Conclusion }

Distortions for the detectors of UVIT have been mapped by imaging a
calibrated grid of holes. The extent of distortions is found to be  0.2\% -
0.4\% of the full field or 3.5$''$ - 7$''$ in the different detectors.  The
corrections for these distortions would make it possible to reduce the
astrometric errors from 3.5$''$ - 7$''$ to 1$''$ or less. 

For verifying accuracy of these measured distortion maps, we have taken
images of few open clusters in the orbit \citep{purni16}. Initial analysis of the VIS
images indicates a marked improvement in the astrometry by application of
corrections for the distortion. The rms error in positions of the stars is
reduced to 2.24 sub-pixel from 6.1 sub-pixel.  A detailed analysis is
under progress \citep{girish16}.

\begin{acknowledgements}

	The authors would like to thank University of Calgary for providing the
	grid, IUCAA for providing the IFOSC images of mask/grid, staff of MGKM
	laboratory for support for IIA images taken with the CPUs and collegues
	of UVIT instrumentation team for all the necessary help.  VG would like
	to thank Dr Annadurai, Director, ISRO Satellite Centre (ISAC) and Dr Anil
	Aggrawal, Group Director, SAG, ISAC and Dr Sreekumar, Director, IIA
	(former GD, SAG, ISAC) for all their encouragement and support.

\end{acknowledgements}


\end{document}